\documentclass[a4paper,10pt,prd,twocolumn,nofootinbib]{revtex4-1}
\usepackage{amssymb,amsfonts,amsmath,graphicx,times,times,amssymb}
\usepackage{soul}
\usepackage{hyperref}
\usepackage{dsfont}

\date{\today}

\begin{document}

\author{David Edward Bruschi}
\affiliation{York Centre for Quantum Technologies, Department of Physics, University of York, YO10 5DD Heslington, UK}
\email{david.edward.bruschi@gmail.com}

\title{Work drives time evolution}

\begin{abstract}
We propose the idea that time evolution of quantum systems is driven by work. The formalism presented here falls within the scope of a recently proposed theory of gravitating quantum matter where extractible work, and not energy, is responsible for gravitation. Our main assumption is that extractible work, and not the Hamiltonian, dictates dynamics. We find that expectation values of meaningful quantities, such as the occupation number, deviate from those predicted by standard quantum mechanics. The scope, applications and validity of this proposal are also discussed.
\end{abstract}

\maketitle


It has been recently proposed that not all energy has a weight \cite{Bruschi:2017}. This novel approach to a theory of gravitation implies a modification of the field equations of gravity, which takes into account the quantum nature of the source. The rationale behind this theory is that, since not all energy can be extracted from quantum systems and converted into work \cite{Pusz:Woronowicz:1978}, only the extractible amount of energy that can be converted into work will gravitate. This proposal has left open many questions, from the renormalisability of the field equations to the correct choice for a time evolution operator compatible with the tenets of the new theory.

Here we propose a time evolution operator for quantum systems that is compatible with this novel theory, i.e., with gravitation. We work in the Heisenberg picture. Given that the source of gravity is not the energy, but the extractible work \cite{Allahverdyan:Balian:2004}, we propose that the Heisenberg equation $\dot{A}=\frac{i}{h}[H,A]+\frac{\partial A}{\partial t}$ for the time evolution of an operator $A$ with the Hamiltonian $H$ should be modified to
\begin{align} \label{new:Heisenberg:equation}
\dot{A}=\frac{i}{\hbar}\left[H,A\right]-\frac{i}{\hbar}[U^\dag_p\,H\,U_p,A]+\frac{\partial A}{\partial t},
\end{align}
where $U_p$ is the \textit{unique} unitary operator\footnote{It acts as the identity in the orthogonal complement of
the subspace spanned by the initial state and the unique passive state.} that maps the initial state $\rho$ to the corresponding passive state $\rho_p$, that is unique up to degeneracies, and $H$ is the Hamiltonian of the system \cite{Horodecki:Oppenheim:2013}. We note here that $U_p$ is a time independent operation.

The right hand side of \eqref{new:Heisenberg:equation} is compatible with the main field equations proposed in recent work aimed at reconciling quantum matter with gravity \cite{Bruschi:2017}. Equation \eqref{new:Heisenberg:equation}, for an operator that does not depend explicitly on time, leads to the formal solution
\begin{align} \label{time:evolution:operator}
U(t)=\overset{\leftarrow}{\mathcal{T}}\exp\left[-\frac{i}{\hbar}\int_0^t\,dt'\,\left(H(t')-U^\dag_p\,H(t')\,U_p\right)\right],
\end{align}
where $\overset{\leftarrow}{\mathcal{T}}$ is the time ordering operator \cite{BandD}.
The operator can be equivalently written as $U(t)=U_H(t)\,\overset{\leftarrow}{\mathcal{T}}\exp\left[\frac{i}{\hbar}\int_0^t\,dt'\,U^\dag_H(t')\,U^\dag_p\,H(t')\,U_p\,U_H(t')\right]$, where we have introduce the standard Heisenberg time evolution operator $U_H(t)$ which reads $U_H(t)=\overset{\leftarrow}{\mathcal{T}}\exp\left[-\frac{i}{\hbar}\int_0^t\,dt'\,H(t')\right]$. This form of the evolution operator allows us to better compare this proposal with the standard approach. 

The formal solution \eqref{time:evolution:operator} does not yield a closed analytical expression in general. However, we can look at simple scenarios where analytical techniques can be applied and an explicit and analytical result can be obtained.

We focus here on systems that have a unique ground state.
Let us start by noting that, if the initial state $|\psi\rangle$ is pure, the operator $U_p$ maps it to the vacuum (ground) state $|0\rangle$ via $U_p|\psi\rangle=|0\rangle$. The operator $U_p$ has the general expression $U_p=\cos\theta\,(|0\rangle\langle0|+|\chi\rangle\langle\chi|)+\sin\theta\,|0\rangle\langle\chi|-\sin\theta\,|\chi\rangle\langle0|+\mathds{1}-|0\rangle\langle0|-|\chi\rangle\langle\chi|$, where we have decomposed the initial state $|\psi\rangle$ as $|\psi\rangle=\cos\theta\,|0\rangle+\sin\theta\,|\chi\rangle$ and $\langle0|\chi\rangle=0$. 

Let us now define the ground state energy $E_0:=H\,|0\rangle$, the energy $E_\chi:=\langle\chi|H|\chi\rangle$ and $\Delta E:=E_\chi-E_0$. We insert the previous expression into \eqref{new:Heisenberg:equation} and obtain
\begin{align} \label{simplified:general:Heisenberg:equation}
\dot{A}=&-\frac{i}{\hbar}\left((2\,(1-\cos\theta)\,E_\chi-\sin\theta^2\,\Delta E)\,[|\chi\rangle\langle\chi|,A]\right.\nonumber\\
&+\sin^2\Delta E\,[|0\rangle\langle0|,A]\nonumber\\
&+\sin\theta\,(E_\chi-\cos\theta\,\Delta E)\,[|\chi\rangle\langle0|,A]\nonumber\\
&+\sin\theta\,(E_\chi-\cos\theta\,\Delta E)\,[|0\rangle\langle\chi|,A]\nonumber\\
&-(1-\cos\theta)\,[H|\chi\rangle\langle\chi|,A]-(1-\cos\theta)\,[|\chi\rangle\langle\chi|H,A]\nonumber\\
&\left.-\sin\theta\,[|0\rangle\langle\chi|H,A]-\sin\theta\,[H|\chi\rangle\langle0|,A]\right).
\end{align}
We note that a special case of \eqref{simplified:general:Heisenberg:equation} occurs when $H\,|\chi\rangle=E\,|\chi\rangle$. 
In this case, the expression \eqref{simplified:general:Heisenberg:equation} reduces to $\dot{A}=-\frac{i}{\hbar}\,\sin\theta\,\Delta E\,(\sin\theta\,[|0\rangle\langle0|,A]-\sin\theta\,[|\chi\rangle\langle\chi|,A]-\cos\theta\,[|\chi\rangle\langle0|,A]-\cos\theta\,[|0\rangle\langle\chi|,A])$.

We now apply our main equation \eqref{simplified:general:Heisenberg:equation} to different classes of states, each of broad interest. We will focus on bosonic states, such as quantum states of light, and we leave it to future work to extend this theory to qubits and fermionic fields. 

We choose initial states $|\psi\rangle$ that are a coherent superposition of the vacuum and an energy eigenstate of the Hamiltonian, i.e., $|\psi\rangle=\cos\theta\,|0\rangle+\sin\theta\,|\chi\rangle$, where $H\,|\chi\rangle=E\,|\chi\rangle$.  
We find it convenient to introduce the vector of operators $\mathbb{X}(t):=(|0\rangle\langle0|,|0\rangle\langle \chi|,|\chi\rangle\langle0|,|\chi\rangle\langle\chi|)^{Tp}$, were $Tp$ stands for transposition and the operators $X_k$  depend on time. The differential equation \eqref{simplified:general:Heisenberg:equation} simplifies, as argued before and, after some algebra, leads to the expression
\begin{align} \label{vector:differential:equation}
\dot{\mathbb{X}}=-\frac{i\,\Delta E}{\hbar}\,\boldsymbol{M}\,\mathbb{X},
\end{align}
where the $4\times4$ matrix $\boldsymbol{M}$ reads
\begin{align}\label{first:matrix}
\boldsymbol{M}=\sin\theta
\begin{pmatrix}
0 & \cos\theta & -\cos\theta & 0\\
\cos\theta & 2\,\sin\theta & 0 & -\cos\theta\\
 -\cos\theta & 0 & -2\,\sin\theta & \cos\theta\\
 0 & -\cos\theta & \cos\theta & 0
\end{pmatrix}.
\end{align}
The differential equation \eqref{vector:differential:equation} has the solution $\mathbb{X}(t)=\exp[-i\frac{\Delta E}{\hbar}\,\boldsymbol{M}\,t]\,\mathbb{X}(0)$. This allows us to compute the probabilities $p_{|0\rangle}:=\langle\psi|0\rangle\langle0|\psi\rangle=\langle\psi|X_1(t)|\psi\rangle$ and $p_{|\chi\rangle}:=\langle\psi|n\rangle\langle n|\psi\rangle=\langle\psi|X_4(t)|\psi\rangle$ of finding the state at time t in either the states $|0\rangle$ or $|\psi\rangle$ respectively. We have
\begin{align}\label{the:main:result}
p_{|0\rangle}(t)=&\cos^2\theta-\cos^2\theta\,\sin^2\,\left(\sin\theta\,\frac{\Delta E}{\hbar}\,t\right)\nonumber\\
p_{|\chi\rangle}(t)=&\sin^2\theta+\cos^2\theta\,\sin^2\,\left(\sin\theta\,\frac{\Delta E}{\hbar}\,t\right).
\end{align}
We are also able to compute another important quantity, i.e., the probability $p_{|\psi\rangle}(t)$ of finding the state of the system at time $t$ in the \textit{same} state as the initial one. This probability is obtained through the solution $\mathbb{X}(t)$ and expressing $|\psi\rangle\langle\psi|(t)$ as $|\psi\rangle\langle\psi|(t)=\cos^2\theta\,|0\rangle\langle0|(t)+\sin\theta\,\cos\theta\,|0\rangle\langle\psi|(t)+\sin\theta\,\cos\theta\,|\psi\rangle\langle0|(t)+\sin^2\theta\,|\psi\rangle\langle\psi|(t)$. Lengthy algebra allows us to find
\begin{align}\label{the:other:main:result}
p_{|\psi\rangle}(t)=&1-\cos^2\theta\,\sin^2\,\left(\sin\theta\,\frac{\Delta E}{\hbar}\,t\right).
\end{align}
Notice that, contrary to what expected for the same computation in the standard Heisenberg picture, $p_{|\psi\rangle}(t)\neq1$ for all times if $0<\theta<\pi$, i.e., if the state is non-classical.

Let us start by focusing on a single bosonic mode with annihilation and creation operators $a,a^\dag$ that satisfy the canonical commutation relations $[a,a^\dag]=1$. 

We choose a free Hamiltonian $H_0$ for simplicity and as the first initial state $|\psi\rangle$ the state $|\psi\rangle=\cos\theta\,|0\rangle+\sin\theta|n\rangle$, where $H_0\,|n\rangle=n\hbar\,\omega\,|n\rangle$, $|0\rangle$ is the vacuum state and $\omega$ is the frequency of the bosonic mode. We note that for $\theta=\pi/2$ the sate is the energy eigenstate $|\psi\rangle=|n\rangle$, while for $\theta=\pi/4$ the state is the maximally coherent state $|\psi\rangle=\frac{1}{\sqrt{2}}[|0\rangle+|n\rangle]$. The corresponding passive state $\rho_p$ to the state $|\psi\rangle$ is the vacuum $|0\rangle$ and one has $|0\rangle=U_p\,|\psi\rangle$, where it is easy to show that $U_p=|0\rangle\langle n|+|n\rangle\langle0|+\sum_{k\neq0,n}|k\rangle\langle k|$ for $\theta=\pi/2$ and $U_p=\frac{1}{\sqrt{2}}[|0\rangle\langle0|+|0\rangle\langle n|-|n\rangle\langle0|+|n\rangle\langle n|]+\sum_{k\neq0,n}|k\rangle\langle k|$ for $\theta=\pi/4$.

We use \eqref{the:main:result} and the matrix \eqref{first:matrix} to find the probabilities $p_{|0\rangle}(t)$ and $p_{|n\rangle}(t)$ for both cases. We have  $p_{|0\rangle}(t)=0$ and $p_{|n\rangle}(t)=1$ for the state $|\psi\rangle=|n\rangle$ with $\theta=\pi/2$, while  we have 
\begin{align}\label{first:main:result}
p_{|0\rangle}(t)=&\frac{1}{2}\,\left[1-\sin^2\,\left(\frac{(n\,\hbar\omega-E_0)\,t}{\sqrt{2}\,\hbar}\right)\right]\nonumber\\
p_{|n\rangle}(t)=&\frac{1}{2}\,\left[1+\sin^2\,\left(\frac{(n\,\hbar\omega-E_0)\,t}{\sqrt{2}\,\hbar}\right)\right].
\end{align}
for the state with $\theta=\pi/4$, i.e., $|\psi\rangle=\frac{1}{\sqrt{2}}[|0\rangle+|n\rangle]$. Note that $p_{|0\rangle}(t)+p_{|n\rangle}(t)=1$ and $p_{|0\rangle}(0)=p_{|n\rangle}(0)=1/2$ as expected. We can also compute the number expectation value $\langle N\rangle:=\langle a^\dag\,a\rangle_{\rho}$, where $a^\dag\,a=\sum_k\,k\,|k\rangle\langle k|$, and we find $\langle N\rangle=\frac{n}{2}(1+\sin^2\,\left(\frac{(n\,\hbar\omega-E_0)\,t}{\sqrt{2}\,\hbar}\right))$. The standard value predicted by quantum mechanics is $\langle N\rangle=\frac{n}{2}$.

We note that the probabilities \eqref{first:main:result} differ dramatically from the expected values $p_{|0\rangle}=p_{|n\rangle}=1/2$ predicted by quantum mechanics.  The conclusion is that the theory predicts that, on average, it is increasingly easier to find the state $|\psi\rangle$ in the eigenstate $|n\rangle$ rather than in $|0\rangle$. This is in line with the idea that gravitating matter initially found in states with quantum coherence tends to ``decohere''. 
Furthermore, the result \eqref{first:main:result} implies that coherence has a \textit{time dependent} effect on the probability of detecting the particle in one of the two allowed eigenstates of the Hamiltonian. 

We now proceed with signature two-mode states of modes $a$ and $b$ with frequencies $\omega_a$ and $\omega_b$ respectively. The annihilation and creation operators are $a,a^\dag$ and $b,b^\dag$, which define the vacuum $|00\rangle=|0\rangle_a\otimes|0\rangle_b$, where $a\,|0\rangle_a=0$ and $b\,|0\rangle_b=0$. The free Hamiltonian is $H_0=\hbar\,\omega_a\,a^\dag\,a+\hbar\,\omega_b\,b^\dag\,b$. Its eigenstates $|nm\rangle$ are defined by $H_0\,|nm\rangle=\hbar\,(n\,\omega_a+m\,\omega_b)\,|nm\rangle$. Let us assume that $E_0=0$ for simplicity here.

We start by studying a separable state $|\psi\rangle=|nm\rangle$. This case is analogous to the single mode case with initial state $|\psi\rangle=|n\rangle$, and passive state $|00\rangle$. We follow the procedure laid down before and set $\theta=\pi/2$. We compute the probabilities $p_{|0\rangle}(t)$ and $p_{|nm\rangle}(t)$ of finding the state in the states $|00\rangle$ or $|nm\rangle$. We find $p_{|0\rangle}(t)=0$ and $p_{|nm\rangle}(t)=1$, as expected.

The next state to be analysed is the separable state $|\psi\rangle=\frac{1}{2}(|0\rangle+|n\rangle)\otimes(|0\rangle+|m\rangle)$. A more convenient way to write this state is $|\psi\rangle=\frac{1}{2}[|00\rangle+|0m\rangle+|n0\rangle+|nm\rangle]$. The operator $U_p$ for this case reads $U_p=U_{p,a}\otimes U_{p,b}$, where $U_{p,a}=\frac{1}{\sqrt{2}}[|0\rangle\langle0|+|0\rangle\langle n|-|n\rangle\langle0|+|n\rangle\langle n|]$ and $U_{p,b}=\frac{1}{\sqrt{2}}[|0\rangle\langle0|+|0\rangle\langle m|-|m\rangle\langle0|+|m\rangle\langle m|]$, and they act on the \textit{different} Hilbert spaces of modes $a$ and $b$ respectively.
We introduce the vector $\mathbb{X}$, which now has $16$ components, and we can compute the solution to the differential equation \eqref{simplified:general:Heisenberg:equation}. Given the structure of the operator $U_p$, the solution simply reads $\dot{\mathbb{X}}=((-\frac{i\,\omega_a}{2}\boldsymbol{M})\oplus(-\frac{i\,\omega_b}{2}\boldsymbol{M}))\,\mathbb{X}$, where $\boldsymbol{M}$ has the same expression as in \eqref{first:matrix}. This implies that we can easily compute the probabilities $p_{|00\rangle}$, $p_{|n0\rangle}$, $p_{|0m\rangle}$ and $p_{|nm\rangle}$ of finding the state $|\psi\rangle$ in either of the states $|00\rangle$, $|n0\rangle$, $|0m\rangle$ or $|nm\rangle$ respectively. We note that, given the structure of the differential equation for this scenario, we have $p_{|00\rangle}=p_{0,n}\,p_{0,m}$, $p_{|n0\rangle}=p_{0,m}\,p_n$, $p_{|0m\rangle}=p_{0,n}\,p_m$ and $p_{|nm\rangle}=p_n\,p_m$, where $p_{0,n}$, $p_n$ and $p_{0,m}$, $p_m$ are the one-mode probabilities \eqref{first:main:result}.
Explicitly we have
\begin{align}\label{second:main:result}
p_{|00\rangle}(t)=&\frac{1}{4}\,\left[1-\sin^2\,\left(n\,\frac{\omega_a\,t}{\sqrt{2}}\right)\right]\,\left[1-\sin^2\,\left(m\,\frac{\omega_b\,t}{\sqrt{2}}\right)\right]\nonumber\\
p_{|n0\rangle}(t)=&\frac{1}{4}\,\left[1+\sin^2\,\left(n\,\frac{\omega_a\,t}{\sqrt{2}}\right)\right]\,\left[1-\sin^2\,\left(m\,\frac{\omega_b\,t}{\sqrt{2}}\right)\right]\nonumber\\
p_{|0m\rangle}(t)=&\frac{1}{4}\,\left[1-\sin^2\,\left(n\,\frac{\omega_a\,t}{\sqrt{2}}\right)\right]\,\left[1+\sin^2\,\left(m\,\frac{\omega_b\,t}{\sqrt{2}}\right)\right]\nonumber\\
p_{|nm\rangle}(t)=&\frac{1}{4}\,\left[1+\sin^2\,\left(n\,\frac{\omega_a\,t}{\sqrt{2}}\right)\right]\,\left[1+\sin^2\,\left(m\,\frac{\omega_b\,t}{\sqrt{2}}\right)\right].
\end{align}
The probabilities \eqref{second:main:result} add up to unity, as expected. 

The number expectation value $\langle N\rangle$ for this case reads $\langle N\rangle=\frac{n}{2}(1+\sin^2\,\left(n\,\frac{\omega_a\,t}{\sqrt{2}}\right))+\frac{m}{2}(1+\sin^2\,\left(m\,\frac{\omega_b\,t}{\sqrt{2}}\right))$.

One can generalise the results \eqref{second:main:result} to a multimode state of $N$ modes, i.e., $|\psi\rangle=\frac{1}{2}(|0\rangle+|n\rangle)_1\otimes(|0\rangle+|m\rangle)_2\otimes\ldots\otimes(|0\rangle+|m\rangle)_N$. One can easily see that the only probability that can reach unity periodically is $p_{|nm\rangle}(t)$, only if all frequencies satisfy $\frac{m\,\omega_m}{\sqrt{2}}=(2k_m+1)\,\omega$, and $k_m\in\mathbb{N}$. 

We draw here an interesting consequence. Let the state $|\psi\rangle$ contain $N$ systems with different frequencies. Then, the probabilities \eqref{second:main:result} will, in general, be strictly smaller than unity and \textit{not} periodic, in sharp contrast with the standard evolution induced by the Heisenberg equation. Detection of the ``excited'' reduced state $|m_1,m_2,\ldots,m_N\rangle$ will be more likely than what predicted from standard theory.

We proceed with looking at the maximally entangled two-mode state $|\psi\rangle=\frac{1}{\sqrt{2}}[|00\rangle+|nm\rangle]$. It is easy to see that, in this case, $U_p=\frac{1}{\sqrt{2}}[|00\rangle\langle00|+|00\rangle\langle nm|-|nm\rangle\langle00|+|nm\rangle\langle nm|]$. Here, again, we set $\theta=\pi/4$. Therefore, we can immediately obtain the final probabilities $p_{|00\rangle}(t)$ and $p_{|nm\rangle}(t)$, which read
\begin{align}\label{third:main:result}
p_{|00\rangle}(t)=&\frac{1}{2}\,\left[1-\sin^2\,\left(\frac{(n\,\omega_a+m\,\omega_b)\,t}{\sqrt{2}}\right)\right]\nonumber\\
p_{|nm\rangle}(t)=&\frac{1}{2}\,\left[1+\sin^2\,\left(\frac{(n\,\omega_a+m\,\omega_b)\,t}{\sqrt{2}}\right)\right].
\end{align}

As a final application of our main equation we study the case of an initial beam splitter-like (or M00N-like) state of two modes $a$ and $b$, i.e., the state $|\psi\rangle$ where $|\psi\rangle=\cos\phi\,|M0\rangle+\sin\phi\,|0N\rangle$. Here we have $H_0\,|M0\rangle=M\,\hbar\,\omega_a\,|M0\rangle$ and $H_0\,|0N\rangle=N\,\hbar\,\omega_b\,|0N\rangle$. Clearly, $\langle0|\psi\rangle=0$ and we have $\theta=\pi/2$. In this problem, the main control parameter is the energy difference $\Delta E:=N\,\hbar\,\omega_b-M\,\hbar\,\omega_a$. The vector $\mathbb{X}$ has nine elements and it is possible to obtain the main differential equation \eqref{vector:differential:equation}. The final formula is not illuminating and we leave the whole details for a more detailed technical work. The main differential equation does not provide an analytical solution for the general case. Nevertheless, we can use it to look at two cases of great interest for theoretical and experimental science. 

The first case occurs when $\Delta E=0$, i.e., when we have a superposition of two eigenstates of the Hamiltonian with the same energy, and $H_0\,|M0\rangle=E\,|M0\rangle$ and $ H_0\,|0N\rangle=E\,|0N\rangle$. In this case, the differential equations \eqref{vector:differential:equation} decouple into three blocks and we find that the probabilities of detecting the state in vectors $|M0\rangle$ and $|0N\rangle$ are $p_{|M0\rangle}(t)=\cos^2\phi$ and $p_{|0N\rangle}(t)=\sin^2\,\phi$ respecitvely, where $p_{|M0\rangle}(t)+p_{|0N\rangle}(t)=1$ as expected. This result is compatible with the use of standard $N00N$ states in negligible gravitational fields, i.e., if we just set $M=N$, $\omega_a=\omega_b=\omega$ and $\phi=\pi/4$ from the start. 

The second case of interest occurs when we study a $N00N$ state where the modes $a$ and $b$ have slightly different energies. This can be taken into account by considering our $M00N$ state with $M=N$ and by introducing the energy shift $\delta E:=N\,\hbar\,(\omega_b-\omega_a)$, where $\epsilon:=\frac{\delta E}{\hbar\,\omega_a}\ll1$. We see that $\epsilon=\delta\omega/\omega_a$, where $\delta\omega:=\omega_b-\omega_a$. We can then employ perturbation theory to our differential equation of the form \eqref{vector:differential:equation} and show that to first order it reads $\mathbb{X}(t)=\boldsymbol{U}_0(t)(\mathds{1}-i\,\epsilon\,\int_0^t dt'\,\boldsymbol{U}_0^\dag(t')\,\boldsymbol{M}^{(1)}\,\boldsymbol{U}_0(t'))$, where $\boldsymbol{U}_0(t)=\exp[-i\,\boldsymbol{M}^{(0)}\,t]$ and the matrix $\boldsymbol{M}$ can be expanded as $\boldsymbol{M}=\boldsymbol{M}^{(0)}+\boldsymbol{M}^{(1)}\,\epsilon$. This solution has a simple analytical expression which to first order reads
\begin{align}\label{fourth:main:result}
p_{|N0\rangle_a}(t)=&\cos^2\phi-4\,\epsilon\,F(\phi)\,\sin^2\left(\frac{N\,\omega_a\,t}{2}\right)\nonumber\\
p_{|0N\rangle_b}(t)=&\sin^2\phi+4\,\epsilon\,F(\phi)\,\sin^2\left(\frac{N\,\omega_a\,t}{2}\right),
\end{align}
where we have introduced $F(\phi):=\sin^4\phi\cos^2\phi\,(1+2\,\cos^2\phi)$ for ease of presentation and we have $\langle N \rangle=N+\mathcal{O}(\epsilon^2)$.
Notice that the results in \eqref{fourth:main:result} can be easily generalised to $M00N$ states where $\frac{\delta E}{\hbar\,\omega_a}\ll1$ and $\delta E:=\hbar\,(N\,\omega_b-M\,\omega_a)$. It is sufficient to replace $\epsilon$ by $\frac{\delta E}{M\,\hbar\,\omega_a}$. We now apply these results to a case of modern interest.

It is a fundamental endeavour that of understanding the role and effects of entanglement in the presence of a gravitational field. There are recent proposals to employ $N00N$ states in Mach-Zehnder-like experiments where the two different paths are located at two different heights in the gravitational field of the Earth \cite{Peters:Chung:2001}. The occurrence of gravitational effects on the state can potentially be measured and can unveil deviations from standard theories, stimulating novel research.
We can apply our results to such a scenario. We assume that the higher path in the gravitational potential can be modelled by a slightly shifted frequency $\omega(L)$ which reads $\omega(L)=(1-\frac{r_S}{2}\frac{L}{r_E^2})\,\omega_0$. Here $r_s$ is the Schwarzschild radius of the Earth, $r_E$ is the radius of the Earth where the lower path lies and the modes have frequency $\omega_0$ and $L$ is the distance between the two paths. This is the standard gravitational redshift formula to first order in $L/r_E\ll1$.
Our probabilities \eqref{fourth:main:result} for this test scenario, where $\phi=\pi/4$, to first order read
\begin{align}\label{gravitational:main:result}
p_{|N0\rangle}(t)=&\frac{1}{2}\left[1+\frac{r_S}{r_E}\,\frac{L}{r_E}\,\sin^2\left(\frac{N\,\omega_0\,t}{2}\right)\right]\nonumber\\
p_{|0N\rangle}(t)=&\frac{1}{2}\left[1-\frac{r_S}{r_E}\,\frac{L}{r_E}\,\sin^2\left(\frac{N\,\omega_0\,t}{2}\right)\right].
\end{align}
Given that $r_S=10^{-2}$m and $r_E=6.371\times10^{6}$m for the Earth, assuming that $L=10^5$m gives us that $\frac{r_S\,L}{r_E^2}\sim4.2\times10^{-11}\ll1$. This allows us to provide a rough estimate of the order of magnitude of the expected effects.

The conclusion here is that, in a Mach-Zehnder type experiment \cite{Peters:Chung:2001}, our theory predicts that the probability $p_{|N0\rangle}(t)$ of detecting the $N$ excitations in the higher path will be slightly higher than the probability $p_{|0n\rangle}(t)$ of detecting them in the lower path.

Finally, it is possible to compute for the probability \eqref{the:other:main:result} of finding \textit{any} initial state $|\psi\rangle$ that is orthogonal to the vacuum state (i.e., $\theta=\pi/2$, $|\psi\rangle=|\chi\rangle$ and $\langle0|\psi\rangle=0$) in the same initial state at a later time. Surprisingly, this reads $p_{|\psi\rangle}(t)=|\langle\psi|\exp[-\frac{i}{\hbar}H_0\,t]|\psi\rangle|^2$, which is the same probability as computed with the standard Heisenberg equation. We note, however, that it is in general not possible to detect such a state with a single measurement, unless it is an eigenstate of the Hamiltonian. In this sense, it is more interesting to compute the probability of finding such a state in any of the eigenstates of the free Hamiltonian.

We now make a few considerations about the validity and scope of the theory and its results. The new oscillating terms that appear in the probabilities \eqref{the:main:result} and \eqref{the:other:main:result} oscillate with a frequency $\omega_{osc}$ that reads $\omega_{osc}=\sin\theta\,\Delta E/\hbar$. The oscillating terms in the probabilities \eqref{fourth:main:result} oscillate with a frequency $\omega_{osc}=N\,\omega_a$. We can introduce a measure $\mathcal{C}$ of coherence for both single and two mode states considered here, which it is defined as the sum of the off diagonal terms in the density matrix \cite{Baumgratz:Cramer:2014}. For both single and two mode states we have that $\mathcal{C}=|\sin(2\,\theta)|$. In the two-mode case, we can also employ the PPT criterion to detect entanglement \cite{Peres:1996,Horodecki:1997}. We employ the Negativity $\mathcal{N}$ to quantify the entanglement and find that $\mathcal{N}=1/2\,|\sin(2\,\theta)|=1/2\,\mathcal{C}$. From here we see that the oscillating frequency $\omega_o$ is directly related to coherence and entanglement measures, i.e., $\omega_{osc}=\sqrt{1-\sqrt{1-\mathcal{C}^2}}\,\Delta E/(\sqrt{2}\,\hbar)$, and vanishes for vanishing coherence and entanglement.

The results presented above assume that it is possible, in principle, to create a perfect pure state of a system, for example of a photon. This is, however, impossible. Every system will be, to some degree, entangled with the environment. This can be seen as a consequence of the third law of thermodynamics. It has to be taken into account when designing an experiment to test the results of this work.
We also note that we have proposed a time evolution operator \eqref{time:evolution:operator} tailored for bosonic system. It is necessary to extend this formalism to include also the evolution of finite dimensional systems, such as qubits, and of fermionic fields.

Furthermore, in this work we have postulated that the time evolution law of a system must depend on its initial state. The predictions that follow need to be tested against experimental evidence. A simple gedanken-experiment that resembles recent proposals for tests of quantum coherence in the gravitational field has been analysed above. In order for this proposal to fit the theory laid down in \cite{Bruschi:2017}, it has to be extended to quantum fields in curved spacetime \cite{BandD}. Although the main equation \eqref{time:evolution:operator} does not make reference to a particular quantisation scheme (i.e., it does not depend on first quantisation explicitly), it has to be compatible with the tenets of general relativity. We leave this task to future work.

Another important aspect is the the interplay between local observers and global observers, i.e., the local aspect of time evolution and the global aspects of it. An observer, Alice, might be in possess of her local quantum state $\rho_A$, which can be mixed or pure. The theory predicts that the time evolution of the system depends on the \textit{global} state of the Universe. Therefore, it might seem that the time evolution of Alice's state depends on information of other subsystems of the Universe, such as the local state of $\rho_B$ of Bob. However, we note that Alice cannot \textit{a priori} uniquely determine how her state $\rho_A$ will evolve unless she has full information of the global state, i.e., she participated to creating the global state.


Finally, we note the theory needs to be compatible with systems that have no ground states, or have a set of degenerate ground states. In addition, the theory needs to be extended to include predictions for interacting systems, therefore understanding the role of interactions within time evolution of physical systems.

To conclude, we have introduced a novel time evolution equation that is compatible with a recently proposed theory of gravitation of quantum systems \cite{Bruschi:2017}. We have shown that the time evolution of quantum states, and the time dependent expectation values of meaningful physical quantities, is different from the corresponding quantities obtained from the standard Heisenberg equation. In particular, highly non classical states, such as coherent superpositions of states with different energies and entangled states, evolve differently than what predicted by quantum mechanics. These states ``tend to be found'' in the highest energy eigenstates. On the contrary, highly classical states, such as single eigenstates of the Hamiltonian or superpositions of eigenstates of the Hamiltonian with same energies, evolve as expected. Furthermore we have showed that $N00N$ states, that can be used in interferometric Mach-Zehnder-like experiments aimed at testing gravity, where two paths are at different heights in the gravitational field, present surprising asymmetries in the occupation numbers in the two paths. The asymmetry in the probabilities depends on the strength of the gravitational field and on the distance between the paths. This corroborates the claim that gravity tends to force quantum states towards eigenstates of the Hamiltonian. We therefore conclude that physical systems must evolve in time according to their initial state.

\textit{Acknowledgements} -- We thank Marcus Huber, Jorma Louko, Leila Khouri, Ivette Fuentes and Jonathan Oppenheim for useful comments and discussions. 

\bibliography{GravQuanTime}

\begin{thebibliography}{9}%
\makeatletter
\providecommand \@ifxundefined [1]{%
 \@ifx{#1\undefined}
}%
\providecommand \@ifnum [1]{%
 \ifnum #1\expandafter \@firstoftwo
 \else \expandafter \@secondoftwo
 \fi
}%
\providecommand \@ifx [1]{%
 \ifx #1\expandafter \@firstoftwo
 \else \expandafter \@secondoftwo
 \fi
}%
\providecommand \natexlab [1]{#1}%
\providecommand \enquote  [1]{``#1''}%
\providecommand \bibnamefont  [1]{#1}%
\providecommand \bibfnamefont [1]{#1}%
\providecommand \citenamefont [1]{#1}%
\providecommand \href@noop [0]{\@secondoftwo}%
\providecommand \href [0]{\begingroup \@sanitize@url \@href}%
\providecommand \@href[1]{\@@startlink{#1}\@@href}%
\providecommand \@@href[1]{\endgroup#1\@@endlink}%
\providecommand \@sanitize@url [0]{\catcode `\\12\catcode `\$12\catcode
  `\&12\catcode `\#12\catcode `\^12\catcode `\_12\catcode `\%12\relax}%
\providecommand \@@startlink[1]{}%
\providecommand \@@endlink[0]{}%
\providecommand \url  [0]{\begingroup\@sanitize@url \@url }%
\providecommand \@url [1]{\endgroup\@href {#1}{\urlprefix }}%
\providecommand \urlprefix  [0]{URL }%
\providecommand \Eprint [0]{\href }%
\providecommand \doibase [0]{http://dx.doi.org/}%
\providecommand \selectlanguage [0]{\@gobble}%
\providecommand \bibinfo  [0]{\@secondoftwo}%
\providecommand \bibfield  [0]{\@secondoftwo}%
\providecommand \translation [1]{[#1]}%
\providecommand \BibitemOpen [0]{}%
\providecommand \bibitemStop [0]{}%
\providecommand \bibitemNoStop [0]{.\EOS\space}%
\providecommand \EOS [0]{\spacefactor3000\relax}%
\providecommand \BibitemShut  [1]{\csname bibitem#1\endcsname}%
\let\auto@bib@innerbib\@empty
\bibitem [{\citenamefont {Bruschi}(2017)}]{Bruschi:2017}%
  \BibitemOpen
  \bibfield  {author} {\bibinfo {author} {\bibfnamefont {D.~E.}\ \bibnamefont
  {Bruschi}},\ }\href@noop {} {\enquote {\bibinfo {title} {On the gravitational
  nature of energy},}\ } (\bibinfo {year} {2017}),\ \Eprint
  {http://arxiv.org/abs/arXiv:1701.00699} {arXiv:1701.00699} \BibitemShut
  {NoStop}%
\bibitem [{\citenamefont {Pusz}\ and\ \citenamefont
  {Woronowicz}(1978)}]{Pusz:Woronowicz:1978}%
  \BibitemOpen
  \bibfield  {author} {\bibinfo {author} {\bibfnamefont {W.}~\bibnamefont
  {Pusz}}\ and\ \bibinfo {author} {\bibfnamefont {S.~L.}\ \bibnamefont
  {Woronowicz}},\ }\href@noop {} {\bibfield  {journal} {\bibinfo  {journal}
  {Comm. Math. Phys.}\ }\textbf {\bibinfo {volume} {58}},\ \bibinfo {pages}
  {273} (\bibinfo {year} {1978})}\BibitemShut {NoStop}%
\bibitem [{\citenamefont {Allahverdyan}\ \emph {et~al.}(2004)\citenamefont
  {Allahverdyan}, \citenamefont {Balian},\ and\ \citenamefont
  {Nieuwenhuizen}}]{Allahverdyan:Balian:2004}%
  \BibitemOpen
  \bibfield  {author} {\bibinfo {author} {\bibfnamefont {A.~E.}\ \bibnamefont
  {Allahverdyan}}, \bibinfo {author} {\bibfnamefont {R.}~\bibnamefont
  {Balian}}, \ and\ \bibinfo {author} {\bibfnamefont {T.~M.}\ \bibnamefont
  {Nieuwenhuizen}},\ }\href@noop {} {\bibfield  {journal} {\bibinfo  {journal}
  {EPL (Europhysics Letters)}\ }\textbf {\bibinfo {volume} {67}},\ \bibinfo
  {pages} {565} (\bibinfo {year} {2004})}\BibitemShut {NoStop}%
\bibitem [{\citenamefont {Horodecki}\ and\ \citenamefont
  {Oppenheim}(2013)}]{Horodecki:Oppenheim:2013}%
  \BibitemOpen
  \bibfield  {author} {\bibinfo {author} {\bibfnamefont {M.}~\bibnamefont
  {Horodecki}}\ and\ \bibinfo {author} {\bibfnamefont {J.}~\bibnamefont
  {Oppenheim}},\ }\href@noop {} {\bibfield  {journal} {\bibinfo  {journal}
  {Nature Communications}\ }\textbf {\bibinfo {volume} {4}},\ \bibinfo {pages}
  {2059 EP } (\bibinfo {year} {2013})}\BibitemShut {NoStop}%
\bibitem [{\citenamefont {Birrell}\ and\ \citenamefont {Davies}(1984)}]{BandD}%
  \BibitemOpen
  \bibfield  {author} {\bibinfo {author} {\bibfnamefont {N.~D.}\ \bibnamefont
  {Birrell}}\ and\ \bibinfo {author} {\bibfnamefont {P.~C.~W.}\ \bibnamefont
  {Davies}},\ }\href@noop {} {\emph {\bibinfo {title} {Quantum fields in curved
  space}}}\ (\bibinfo  {publisher} {Cambridge University press},\ \bibinfo
  {year} {1984})\BibitemShut {NoStop}%
\bibitem [{\citenamefont {Peters}\ \emph {et~al.}(2001)\citenamefont {Peters},
  \citenamefont {Chung},\ and\ \citenamefont {Chu}}]{Peters:Chung:2001}%
  \BibitemOpen
  \bibfield  {author} {\bibinfo {author} {\bibfnamefont {A.}~\bibnamefont
  {Peters}}, \bibinfo {author} {\bibfnamefont {K.~Y.}\ \bibnamefont {Chung}}, \
  and\ \bibinfo {author} {\bibfnamefont {S.}~\bibnamefont {Chu}},\ }\href
  {http://stacks.iop.org/0026-1394/38/i=1/a=4} {\bibfield  {journal} {\bibinfo
  {journal} {Metrologia}\ }\textbf {\bibinfo {volume} {38}},\ \bibinfo {pages}
  {25} (\bibinfo {year} {2001})}\BibitemShut {NoStop}%
\bibitem [{\citenamefont {Baumgratz}\ \emph {et~al.}(2014)\citenamefont
  {Baumgratz}, \citenamefont {Cramer},\ and\ \citenamefont
  {Plenio}}]{Baumgratz:Cramer:2014}%
  \BibitemOpen
  \bibfield  {author} {\bibinfo {author} {\bibfnamefont {T.}~\bibnamefont
  {Baumgratz}}, \bibinfo {author} {\bibfnamefont {M.}~\bibnamefont {Cramer}}, \
  and\ \bibinfo {author} {\bibfnamefont {M.~B.}\ \bibnamefont {Plenio}},\
  }\href@noop {} {\bibfield  {journal} {\bibinfo  {journal} {Phys. Rev. Lett.}\
  }\textbf {\bibinfo {volume} {113}},\ \bibinfo {pages} {140401} (\bibinfo
  {year} {2014})}\BibitemShut {NoStop}%
\bibitem [{\citenamefont {Peres}(1996)}]{Peres:1996}%
  \BibitemOpen
  \bibfield  {author} {\bibinfo {author} {\bibfnamefont {A.}~\bibnamefont
  {Peres}},\ }\href@noop {} {\bibfield  {journal} {\bibinfo  {journal} {Phys.
  Rev. Lett.}\ }\textbf {\bibinfo {volume} {77}},\ \bibinfo {pages} {1413}
  (\bibinfo {year} {1996})}\BibitemShut {NoStop}%
\bibitem [{\citenamefont {Horodecki}(1997)}]{Horodecki:1997}%
  \BibitemOpen
  \bibfield  {author} {\bibinfo {author} {\bibfnamefont {P.}~\bibnamefont
  {Horodecki}},\ }\href {\doibase
  http://dx.doi.org/10.1016/S0375-9601(97)00416-7} {\bibfield  {journal}
  {\bibinfo  {journal} {Physics Letters A}\ }\textbf {\bibinfo {volume}
  {232}},\ \bibinfo {pages} {333 } (\bibinfo {year} {1997})}\BibitemShut
  {NoStop}%
\end{thebibliography}%

\end{document}